\documentclass{eas}
\usepackage{graphicx}
\begin{document}

\TitreGlobal{Mass Profiles and Shapes of Cosmological Structures}

\title{Contraction of Dark Matter Halos \\ in Response to Condensation of Baryons}

\author{Oleg Y. Gnedin}
\address{Ohio State University,
         Department of Astronomy;
         \mbox{\tt ognedin@astronomy.ohio-state.edu}}

\runningtitle{Adiabatic Contraction of Dark Matter}

\setcounter{page}{1}

\index{Oleg Y. Gnedin}

\begin{abstract}
  The cooling of baryons in the centers of dark matter halos leads to a
  more concentrated dark matter distribution.  This effect has
  traditionally been calculated using the model of adiabatic
  contraction, which assumes spherical symmetry, while in hierarchical
  formation scenarios halos grow via multiple violent mergers.  We
  test the adiabatic contraction model in high-resolution cosmological
  simulations and find that the dissipation of gas indeed increases
  the density of dark matter and steepens its radial profile compared
  to the case without cooling.  Although the standard model
  systematically overpredicts the increase of dark matter density, a
  simple modification of the assumed invariant from $M(r)r$ to
  $M(\bar{r})r$, where $\bar{r}$ is the orbit-averaged particle
  position, reproduces the simulated profiles within 10\%.
\end{abstract}

\maketitle

\section{Testing the Adiabatic Contraction Model}

In hierarchical galaxy formation models, stars are formed in
condensations of cooled baryons within dark matter halos.  The
response of dark matter to baryonic infall has traditionally been
calculated using the model of adiabatic contraction (AC).  The
standard model of AC, introduced by Blumenthal et al. (1986), assumes
spherical symmetry, homologous contraction, circular particle orbits,
and conservation of the angular momentum: $M(r)r = {\rm const}$, where
$M(r)$ is the total mass enclosed within radius $r$.  Because of the
simplicity of this model, it has been routinely applied in the
modeling of galaxies and clusters of galaxies.

It is amazing, however, that the model of AC has not been tested in a
cosmological context.  The hierarchical formation scenario is
considerably more complex than the simple picture of quiescent cooling
in a static spherical halo.  Each halo is assembled via a series of
mergers of smaller halos, in which the cooling of gas and contraction
of dark matter occur separately.  The gas can be re-heated by shocks
and star formation feedback, and in addition early-type galaxies may
undergo dissipationless mergers after the gas is exhausted.

In order to test the AC model, we have analyzed high-resolution
cosmological simulations of eight cluster-sized and one galaxy-sized
systems in a concordance $\Lambda$CDM model (Gnedin et al. 2004).  The
simulations are performed with the Adaptive Refinement Tree (ART)
$N$-body$+$gasdynamics code (Kravtsov 1999).  For each halo, we
analyze two sets of simulations which start from the same initial
conditions but include different physical processes.  The first set of
simulations follows the dynamics of gas ``adiabatically'',
i.e. without radiative cooling.  The second set of simulations includes
gas cooling and heating, and star formation.  The difference in the
halo profiles between the two runs is therefore due to the
condensation of baryons.

\begin{figure}[t]
\centerline{\includegraphics[width=3.5truein]{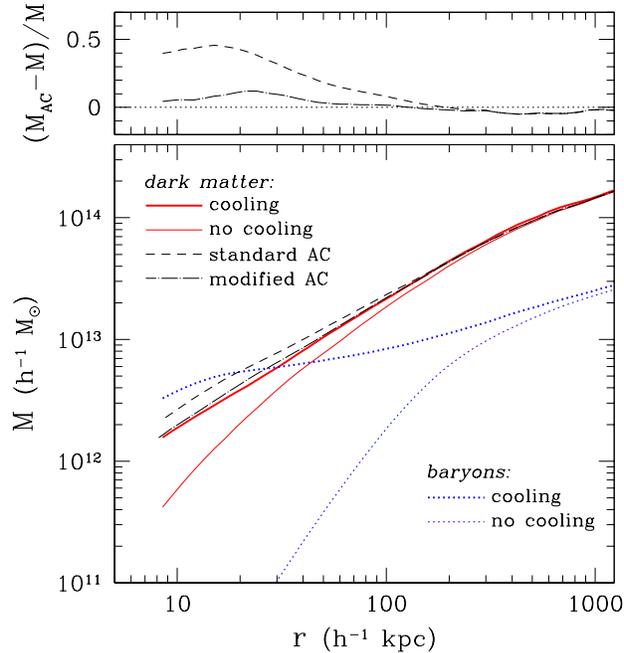}}
\caption{Enclosed mass profiles of dark matter ({\it solid}) and
  baryons (gas+stars, {\it dotted}) in the adiabatic ({\it thin}) and
  cooling ({\it thick}) runs in a simulated cluster of galaxies.  {\it Dashed}
  curve shows the prediction of the standard AC model, while {\it
  dot-dashed} curve shows the improved model.  The profiles are
  truncated at four resolution elements of the simulation.
  {\it Top panel:} relative mass difference between the dark matter
  profile in the cooling run and the prediction of the standard AC
  model ({\it dashed}) or our modified model ({\it dot-dashed}).
  \label{fig:cl6_m}}
\end{figure}

\begin{figure}[t]
\centerline{\includegraphics[width=3.2truein]{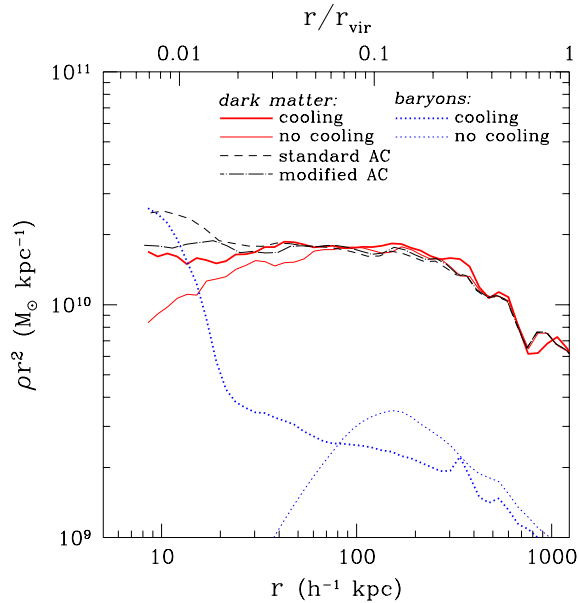}}
\caption{Density profile of the cluster shown in
  Fig. \protect\ref{fig:cl6_m}, with the same line types.
  We plot the combination $\rho(r)r^2$ in order to
  emphasize the differences at small radii.
  \label{fig:cl6_den}}
\end{figure}

Figures \ref{fig:cl6_m} and \ref{fig:cl6_den} show that cooling leads
to an increase in the dark matter density in a representative cluster
of galaxies within $r < 50 h^{-1}$ kpc or $r/r_{\rm vir} < 0.04$.  The
difference in the enclosed mass increases substantially with
decreasing radius.  Note that although the simulations may suffer from
the ``overcooling of baryons'', we can still test whether the AC model
works.  In fact, the larger effect of cooling allows us to emphasize
the difference between the simulations and the model.  The standard
model of AC predicts the overall mass enhancement but systematically
overestimates its magnitude in the inner regions.  This effect has
already been noticed in the original study of Blumenthal et
al. (1986).

\begin{figure}[t]
\centerline{\includegraphics[width=3.2truein]{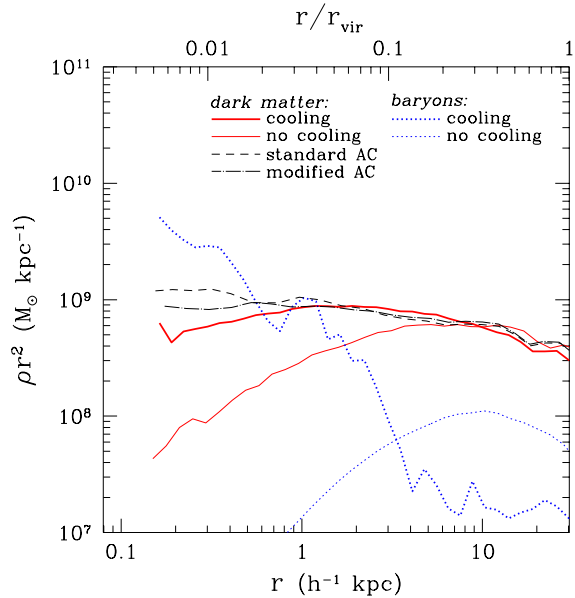}}
\caption{Density profile in the galaxy-sized halo at $z=4$.
  Lines types are as in Fig. \protect\ref{fig:cl6_m}.
  \label{fig:gal_den}}
\end{figure}

In the galaxy-sized halo at $z=4$ (Fig. \ref{fig:gal_den}) the dark
matter density is enhanced within a larger fraction of the virial
radius, $r/r_{\rm vir} < 0.1$, than in the cluster runs.  It is due to
a considerably higher fraction of cold gas in the galaxy run (80\%
vs. $20-30\%$ for clusters), as well as a different mix of gas and
stars (most of the baryon mass within the central regions of clusters
is in the stellar component of the cD galaxy, while the galaxy at
$z=4$ is very gas-rich).  Because of the differences in the gas
densities, temperatures, and cooling times, the cluster and galaxy
simulations probe qualitatively different regimes of central baryon
condensation.  In both regimes the AC model works surprisingly well.

Nevertheless, many of the assumptions of the standard AC model are not
valid.  The orbits of particles in realistic dark matter halos are
highly eccentric.  For such orbits the combination $M(r)r$, which
forms the basis of the standard AC model, varies with the orbital
phase and is not adiabatically invariant.  A better version of AC
model can be constructed by taking into account eccentricities of
particle orbits.  An exact analytic form of the adiabatic invariant is
not known in general case, but Sellwood \& McGaugh (2005) suggest a
numerical solution based on remapping the distribution function.
However, this exact solution to the spherical collapse problem with
conserved orbital actions may not necessarily apply in the case of
hierarchical merging, where the action variables do change.  Instead,
we can look for a convenient approximation to the realistic evolution
of halos in the cosmological simulations.

We proposed a modified adiabatic contraction model (Gnedin et
al. 2004) based on conservation of the product of the current radius
of a spherical shell, $r$, and the mass enclosed within the
orbit-averaged radius, $\bar{r}$:
$$
  M(\bar{r}) \, r = {\rm const}.
$$
The distribution of particle orbits with the energies and angular
momenta found in cosmological simulations is close to isotropic, with
a wide range of eccentricities.  We found that in the range $10^{-3} <
r/r_{\rm vir} < 1$, the orbit-averaged radius for particles at a
current radius $r$ can be described by a power law relation
$$
  {\bar{x}} = A \, x^{w}, \quad x\equiv r/r_{\rm vir},
$$
with small variations in the parameters $A\approx 0.85 \pm 0.05$ and
$w \approx 0.8 \pm 0.02$ from halo to halo and from epoch to epoch.
These two equations allow one to calculate the final profile of a
hierarchically-assembled dark matter halo.

Figures \ref{fig:cl6_m}--\ref{fig:gal_den} show that the modified
model provides a more accurate description of the simulation results
than the standard model.  Although there are still differences between
the model and simulation profiles, the model does not systematic over-
or under-predict the mass for most of the objects.  Typical mass
deviations are less than 10\%.

Given that dark matter halos assemble via mergers and violent
relaxation, it is somewhat surprising that the AC model reproduces the
simulation results so well.  The success of the model seems to imply
that the effect of central baryon condensation on the dark matter
distribution is independent of the way in which this condensation is
assembled.  It may also be due to the fact that the central region is
dominated by particles from a single densest progenitor.  If the
progenitor halo contracts in response to the cooling of baryons early
on and then approximately preserves the shape of its inner density
profile during subsequent mergers, as suggested by collisionless
merger simulations discussed next, the AC model applied to the final
mass distribution would work.

\section{How steep is the central dark matter profile?}

Our results have important implications for the efforts to test
predictions of the CDM model observationally.  The test that received
much attention recently is the density distribution in the inner
regions of galaxies and clusters.  Collisionless simulations indicate
that the dark matter density profile in the absence of gas dissipation
develops an inner cusp, $\rho(r) \propto r^{-\gamma}$.  Using our AC
model, we can analytically predict the change of the asymptotic inner
slope $\gamma$ due to the condensation of baryons.  If the final
baryon density at $r/r_{\rm vir} \ll 1$ can also be written as a power
law, $\rho_{\rm b}(r) \propto r^{-\nu}$, then the post-contraction
dark matter density is
$$
  \rho(r) \propto r^{-\gamma'}, \quad 
  \gamma' = {\gamma + (3-\gamma)w\nu \over 1 + (3-\gamma)w}.
$$
For $\nu=1$, which corresponds to a Hernquist profile or an
exponential disk, the asymptotic slope of an NFW profile ($\gamma =
1$) after the contraction remains the same, $\gamma' = 1$.  If the
baryons are more concentrated, $\nu > \gamma$, the final slope becomes
steeper than the initial slope, $\gamma' > \gamma$.  For an isothermal
sphere, $\nu \approx 2$ (and $w \approx 1$ in our model) the inner
slope of the dark matter profile will be close to the slope of the
baryon profile (Gnedin et al. 2004).

For realistic cases, the profiles of dark matter and baryons on scales
$r \sim 0.01 \, r_{\rm vir}$ should be quite similar.  These scales
are exactly where the mass profiles of massive ellipticals and central
cluster galaxies are probed by spectroscopic measurements and
gravitational lensing (e.g., Treu \& Koopmans 2004, Sand et al. 2004)
and attempts are made to infer the distribution of dark matter in
these systems by subtracting the contribution of baryons from the
total profile.  Since stars dominate the mass in the inner regions,
only a small over-estimate of the stellar mass-to-light ratio may lead
to a significant under-estimate of the residual dark matter density.
Better results can be obtained by using the observed baryon
distribution to calculate the contracted dark matter profile and
fitting the sum of the two to the total measured mass profile.

An important hypothesis has been put forth recently by Loeb \& Peebles
(2003) and Gao et al. (2004) that the NFW profile is a dynamical
attractor, in the sense that remnants of dissipationless mergers are
driven to a profile with the central cusp $\rho(r)\propto r^{-1}$,
even if their progenitors had steeper profiles.  Under this
hypothesis, after an early epoch of star formation in elliptical
galaxies, subsequent dissipationless mergers would erase the
cooling-induced central concentration of dark matter.  Together with
observation that stars dominate in the centers of galaxies, it implies
that the dark matter profile in galaxies which experienced
dissipationless mergers is shallower than $r^{-1}$.

Our results do not support the attractor hypothesis.  In the cluster
runs with cooling and star formation, the growth of the cD galaxy
leads to a significant increase of the dark matter mass enclosed
within the inner $10\, h^{-1}$ kpc (four times that in the adiabatic
runs).  The dark matter density increases, not decreases, at lower
redshifts.  Even when we assume, as Gao et al., that star formation
and gas cooling in massive ellipticals effectively stop at $z \approx
2$ and rerun the simulation with subsequent evolution being purely
dissipationless, we still find that the dark matter distribution is
steeper than in the adiabatic case (but less so than in the original
run where cooling continued until $z=0$).  Collisionless mergers that
the main cluster progenitor has undergone since $z=2$ apparently have not
erased the steepening of the profile due to the cooling.

Controlled experiments by Boylan-Kolchin \& Ma (2004) and Kazantzidis
et al. (2004) also show that cuspy density profiles are preserved in
dissipationless mergers.  The merger remnant of two identical halos
with cuspy profiles retains the initial inner slope, while mergers of
the cored and cuspy halos produce a cuspy remnant.  Their analysis
shows that the density cusp is remarkably stable and the merger
remnants retain the memory of the density profiles of their
progenitors.  Thus the effects of baryon dissipation in the
progenitors are retained, at least partially, in the density
distribution of their descendant.

\bigskip

{\sc Contra} is a publicly available code that calculates the
contraction of a dark matter halo in response to condensation of
baryons in its center.  It is available for download at
{\tt http://www.astronomy.ohio-state.edu/$\sim$ognedin/contra/}

\end{document}